\documentclass[12pt,twocolumn,showpacs]{iopart}
\usepackage{iopams}
\usepackage{epsfig}
\usepackage{graphicx}
\usepackage[usenames]{color}
\usepackage[normalem]{ulem}
\usepackage{soul}
\usepackage{subfigure}
\usepackage{float}

\begin{document}

\title{Solitons under spatially localized cubic-quintic-septimal
nonlinearities}
\author{H. Fabrelli$^{1}$, J. B. Sudharsan$^{2}$, R. Radha$^{2}$, A. Gammal$%
^{1}$, Boris A. Malomed$^{3}$}
\address{$^{1}$Instituto de F\'isica, Universidade de S\~ao Paulo, 05508-090, S\~ao Paulo,Brazil\\$^{2}$Center for Nonlinear Science (CeNSc), PG and Research Department of Physics, Government College for Women (Autonomous), Kumbakonam 612001, Tamil Nadu, India\\$^{3}$Department of Physical Electronics, School of Electrical Engineering, Tel
Aviv University, Tel Aviv 69978, Israel
and Laboratory of Nonlinear-Optical Informatics, ITMO University, St.
Petersburg 197101, Russia} \ead{vittal.cnls@gmail.com$ ^2$}

\begin{abstract}
We explore stability regions for solitons in the nonlinear Schr\"{o}dinger
equation with a spatially confined region carrying a combination of
self-focusing cubic and septimal terms, with a quintic one of either
focusing or defocusing sign. This setting can be implemented in optical
waveguides based on colloids of nanoparticles. The solitons' stability is
identified by solving linearized equations for small perturbations, and is
found to fully comply with the Vakhitov-Kolokolov criterion. In the limit
case of tight confinement of the nonlinearity, results are obtained in an
analytical form, approximating the confinement profile by a delta-function.
It is found that the confinement greatly increases the largest total power
of stable solitons, in the case when the quintic term is defocusing, which
suggests a possibility to create tightly confined high-power light beams
guided \ by the spatial modulation of the local nonlinearity strength.
\end{abstract}

\pacs{03.75.Lm; 05.45.Yv; 42.65.Tg}
\maketitle

\section{Introduction}

The importance of spatial solitons, which are maintained by the balance
between diffraction and self-focusing, in modern photonics is commonly known
\cite{buryak}-\cite{zchensegev}. In particular, the ubiquitous Kerr
self-focusing nonlinearity is modeled by the cubic nonlinear Schr\"{o}dinger
equation (NLSE), which predicts stable solitons in optical fibers and planar
waveguides, as well as in many other systems \cite{Agrawalbook}. On the
other hand, more complex effects, such as the generation of higher-order
harmonics \cite{harmonics}, filamentation \cite{filamentation}, saturation
of the Kerr nonlinearity \cite{saturation}, modulational instability in
metamaterials \cite{meta}, creation of quasi-stable optical beams with
embedded vorticity \cite{quasi-stable}, and other nonlinear phenomena make
it necessary to add higher-order nonlinearities to the NLSE. In particular,
the recently reported stable propagation of (2+1)-dimensional fundamental
spatial solitons in bulk dielectric media is explained by the action of the
cubic-quintic (CQ) \cite{2+1} or quintic-septimal \cite{quintic-septimal}
nonlinearities, in which the sign of the higher term is defocusing. It was
also predicted that the CQ nonlinearity is sufficient to support stable two-
\cite{Michinel} and three- \cite{nine authors} dimensional (2D and 3D)
solitons with embedded vorticity.

Recently, it has been demonstrated that various combinations of optical
nonlinearities, including CQ and septimal terms of either sign, can be
engineered in colloids composed of metallic nanoparticles \cite{colloid}.
This finding has suggested a detailed analysis of 1D solitons under the
action of the combined cubic-quintic-septimal (CQS) nonlinearity \cite%
{reyna4}, the most essential issue being stability of such soliton families
for different combinations of signs in front of the CQS terms. In
particular, a noteworthy result is that, in spite of the well-known fact
that the focusing quintic and septimal terms give rise, respectively, to the
critical and supercritical collapse in the 1D NLSE \cite{berge,Juul}, the 1D
equation with the focusing signs of all the nonlinear terms in the CQS
combination maintains stable soliton families (similar to the previously
known fact that the combination of self-attractive cubic and quintic terms
also generates stable solitons \cite{PhysicaD}).

Another direction of the work with nonlinearity, which has also drawn much
attention in the course of the last decade, is the use of spatially
modulated nonlinear terms for the creation and stabilization of various
species of solitons \cite{Barcelona2}. In particular, a stepwise radial
change of the local strength of the cubic self-focusing is sufficient to
stabilize Townes solitons in the 2D space \cite{HS}. It has also been
recently observed that the spatially modulated nonlinearity could also be
contributed to earlier onset of Faraday and resonant waves \cite%
{scalarFW,binaryFW}. It was also demonstrated that stable 2D solitons with
embedded higher-order vorticity, up to $S=6$, can be supported by localized
cubic self-focusing in a combination with the harmonic-oscillator trapping
potential \cite{we}, while previously no stable states with $S\geq 2$ could
be constructed using local nonlinearity and trapping potentials \cite%
{review,Barcelona2}.

The above-mentioned findings suggest to seek for possibilities to predict
stable solitons supported by the CQS nonlinearity subject to the spatial
modulation, which is the objective of the present work. In effectively
planar waveguides built of the above-mentioned colloidal materials, which
provide the CQS nonlinearity for the creation of spatial solitons, the
modulation can be induced by making the waveguide's thickness a function of
the propagation distance. Higher-order nonlinearities may also be provided
by resonant dopants (see, e.g., Ref. \cite{Scully}), in which case the
spatial modulation is naturally imposed by inhomogeneity of the dopant
density along the propagation distance. We focus on the most interesting
\textquotedblleft sandwich" setting, when the lowest and highest cubic and
septimal terms are self-focusing, while the intermediate quintic one may
have either sign. A challenge is to secure stability of the solitons when,
in particular, the competition between focusing and defocusing (quintic)
terms may support stable solitons, in spite of the possibility of the onset
of the supercritical collapse, driven by the septimal self-focusing term.
The model, based on an appropriate version of the NLSE, is introduced in
Section II. Accurate results of a systematic numerical analysis, focused on
the solitons' stability, are reported in Section III. For the limiting case
of tight confinement of the nonlinearity, approximate analytical results,
which are found to be in good agreement with the numerical counterparts, and
thus help to explain the numerical findings, are presented in Section IV.
The paper is concluded in Section V.

\section{The model}

The 1D NLSE for amplitude $\Psi \left( x,z\right) $ of the electromagnetic
wave with spatially-localized CQS nonlinear terms is written in the scaled
form for the spatial domain as

\begin{equation}
i\frac{\partial \Psi }{\partial z}+\frac{1}{2}\frac{\partial ^{2}\Psi }{%
\partial x^{2}}+\left[ G_{3}(x)P|\Psi |^{2}+G_{5}(x)P^{2}|\Psi |^{4}\right.
\left. +G_{7}(x)P^{3}|\Psi |^{6}\right] \Psi =0,  \label{time_dep}
\end{equation}%
with normalization condition
\begin{equation}
\int_{-\infty }^{+\infty }|\Psi (x,z)|^{2}dx=1  \label{1}
\end{equation}%
and total power $P$. Here $\ z$ and $x$ are the propagation distance and
transverse coordinate, and $x$-dependent coefficients in front of the cubic,
quintic, and septimal ($n=3,5,7$, respectively) terms are defined as
\begin{equation}
G_{n}(x)=g_{n}\exp \left( -b^{2}x^{2}/2\right) ,  \label{gaussian}
\end{equation}%
where $b$ determines the size of the nonlinearity-bearing region, $%
|x|~\lesssim 1/b$. Further, being interested, as said above, in the case
when both the cubic and septimal terms are self-focusing, we use the
possibility to rescale the variables and parameters in Eqs. (\ref{time_dep})
and (\ref{gaussian}), as%
\begin{eqnarray}
\nonumber
z &\equiv &\sqrt{g_{7}/g_{3}^{3}}z^{\prime },~x\equiv \left(
g_{7}/g_{3}^{3}\right) ^{1/4}x^{\prime },~\Psi \equiv \left(
g_{3}^{3}/g_{7}\right) ^{1/8}\Psi ^{\prime },   \\
~P &\equiv &\left( g_{3}g_{7}\right) ^{-1/4}P^{\prime },~b\equiv \left(
g_{3}^{3}/g_{7}\right) ^{1/4}b^{\prime },  \label{scale}
\end{eqnarray}%
so as to fix the accordingly transformed cubic and septimal coefficients
(primes appearing in Eq. (\ref{scale}) are dropped below),
\begin{equation}
g_{3}=g_{7}=1,  \label{=1}
\end{equation}%
while $b$, $g_{5}$ and $P$ are kept as irreducible control parameters, $%
g_{5}>0$ and $<0$ corresponding, severally, to the focusing and defocusing
quintic terms. Of course, the normalization conditions (\ref{=1}) may be
chosen in a different from, if that may be more convenient for some
purposes.

The dynamical model based on Eq. (\ref{time_dep}) with $g_{5}>0$ can be also
derived as an approximate form of the Gross-Pitaevskii equation (GPE) for a
relatively dense atomic Bose-Einstein condensate with attractive
inter-atomic interactions. Starting from the 3D GPE with the cubic
self-attractive nonlinearity, the reduction of the dimension from 3D to 1D
for a condensate tightly trapped in a cigar-shaped potential leads to the
NLSE for the 1D mean-field wave function $\Psi \left( x,t\right) $ with
nonpolynomial nonlinearity $\sim \left[ 1-(3/2)g|\Psi |^{2}\right] (1-g|\Psi
|^{2})^{-1/2}\Psi ,$ where $g>0$ is an effective self-attraction coefficient
\cite{Luca}. The truncated expansion of this expression in powers of $|\Psi
|^{2}$ up to the third order leads to the NLSE in the form of Eq. (\ref%
{time_dep}), with $G_{5}\sim g^{2}>0$, and $z$ replaced by scaled time $t$.
The spatial modulation of the nonlinearity coefficient, $g(x)$, can be
imposed by means of the experimentally available method based on the
Feshbach resonance controlled by a spatially inhomogeneous optical \cite{opt}
or magnetic \cite{we} field.

Stationary states with propagation constant $k$ are looked for as solutions
to Eq. (\ref{time_dep}) in the form of $\Psi (x,z)=\psi (x)e^{ikz}$, with
real function $\psi (x)$ satisfying the ordinary differential equation,

\begin{equation}
k\psi =\frac{1}{2}\frac{d^{2}\psi }{dx^{2}}+\exp \left( -\frac{1}{2}%
b^{2}x^{2}\right) P\psi ^{3}\left( 1+g_{5}P\psi ^{2}+P^{2}\psi ^{4}\right) .
\label{stationary}
\end{equation}
Stability of the stationary states is addressed by taking perturbed
solutions as $\Psi (x,t)=\left[ \psi (x)+\delta \psi (x,z)\right] \exp
\left( ikz\right) $, where small perturbation $\delta \psi (x,t)$ evolves
according to the linearized equation,

\begin{eqnarray}
\nonumber
\left( i\frac{\partial }{\partial z}-k+\frac{1}{2}\frac{\partial ^{2}}{%
\partial x^{2}}\right) \delta \psi +\exp \left( -\frac{1}{2}%
b^{2}x^{2}\right) P\psi ^{2}(x)\times \lbrack (\delta \psi ^{\ast }+2\delta
\psi )+  \\
g_{5}P\psi ^{2}(x)(2\delta \psi ^{\ast }+3\delta \psi )+P^{2}\psi
^{4}(x)(3\delta \psi ^{\ast }+4\delta \psi )]=0,  \label{psi}
\end{eqnarray}%
\newline
with $\ast $ standing for the complex conjugate.\ For perturbation
eigenfunctions sought for in the usual form \cite%
{Yang,pushkarov,skarka,yamazaki}, $\delta \psi (x,z)=u(x)\exp \left(
i\lambda z\right) +v^{\ast }(x)\exp \left( -i\lambda ^{\ast }z\right) $, the
corresponding eigenvalue problem for $\lambda $ amounts to the system of
coupled linear equations:

\begin{eqnarray}
\nonumber
\frac{1}{2}\frac{d^{2}u}{dx^{2}}+\exp \left( -\frac{1}{2}b^{2}x^{2}\right)
P\psi ^{2}(x)\left[ (2u+v)\right.  \\
\nonumber
\left. +g_{5}P\psi ^{2}(x)(3u+2v)+P^{2}\psi ^{4}(x)(4u+3v)\right]
-ku=\lambda u,   \\
\nonumber
\frac{1}{2}\frac{d^{2}v}{dx^{2}}+\exp \left( -\frac{1}{2}b^{2}x^{2}\right)
P\psi ^{2}(x)\left[ (2v+u)\right.   \\
\left. +g_{5}P\psi ^{2}(x)(3v+2u)+P^{2}\psi ^{4}(x)(4v+3u)\right]
-kv=-\lambda v.  \label{bdgsis3d}
\end{eqnarray}%
As usual, the stability condition is that all eigenvalues $\lambda $
(eigenfrequencies of the small-perturbation modes) must be purely real.

Stationary equation (\ref{stationary}) was solved numerically by means of a
relaxation algorithm, which is outlined in Ref. \cite{brtka}, starting with
spatial stepsize $\Delta x=0.01$, and reducing it for solutions with a
shrinking size. For the verification of the results, stationary solutions
were also produced, for $g_{5}>0$ and $b>2$, with the help of a shooting
Runge-Kutta algorithm, as described in Ref. \cite{gammal1999}. The time
evolution governed by Eq. (\ref{time_dep}) was simulated using a
Crank-Nicolson algorithm \cite{pmadikari}, with spatial stepsizes similar to
the above-mentioned ones, employed for the realization of the relaxation
method, and temporal step $\Delta t=0.001$.

The linear system (\ref{bdgsis3d}) was solved, using $\psi ^{2}(x)$ produced
by the relaxation method, and putting the system on a spatial grid composed
of up to $1000$ points. For carrying out the solution procedure, the
corresponding matrix was diagonalized using a LAPACK routine \cite{LAPACK},
that eventually provided all the eigenvalues.

\section{Numerical results}

\begin{figure}[h]
\centering
\subfigure[]{\includegraphics[scale=0.7325]{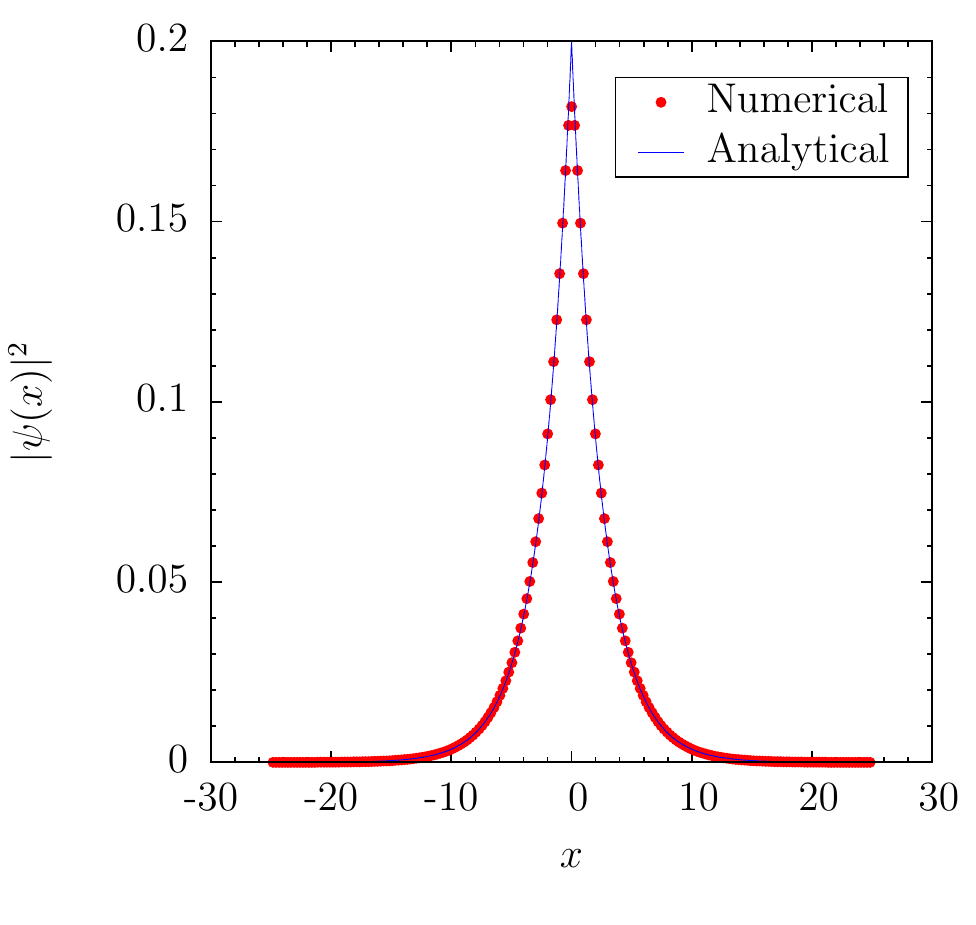}} \subfigure[]{%
\includegraphics[scale=0.7]{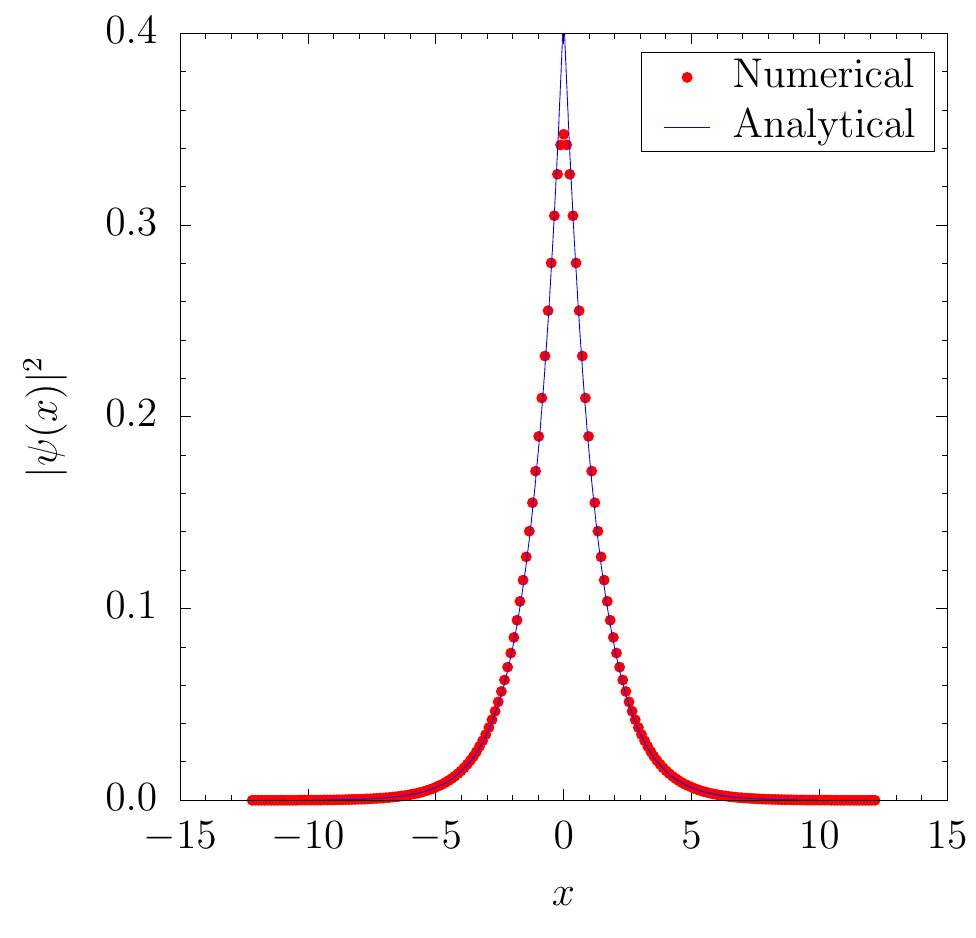}} \subfigure[]{%
\includegraphics[scale=0.7]{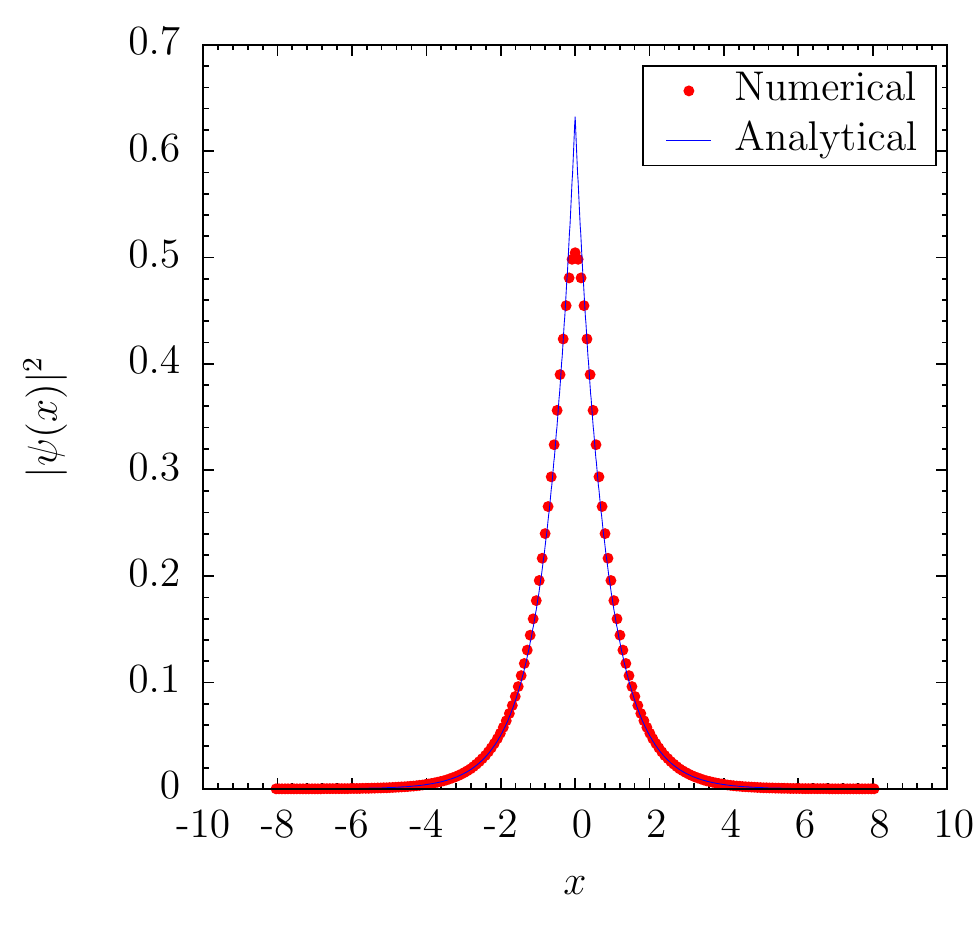}}
\caption{Chains of dots show solitons produced by the numerical solution of
Eq. (\protect\ref{stationary}) with $b=3$ and $g_{5}=-1$, along with
solutions predicted by the analytical approximation based on Eqs. (\protect
\ref{lambda}) and (\protect\ref{quadr}). The propagation constants of the
solutions are (a) $k=0.02$, $P=1.89$ (a stable solution), (b) $k=0.085$, $%
P=P_{\max }=2.22$ (at the stability boundary, see Fig. \protect\ref{mu_n}
below), and (c) $k=0.2$, $P=2.13$ (an unstable solution).}
\label{examples}
\end{figure}

Typical examples of numerically found solitons, along with their
counterparts predicted by the analytical approximation [see Eqs. (\ref%
{lambda}) and (\ref{quadr}) below], are displayed in Fig. \ref{examples}.
The shape of the solitons supported by the confined nonlinearity is close to
that of \textit{peakons}, which are usually produced by models with
nonlinear dispersion, such as the Camassa-Holm equation \cite{CH} and the
continual limit of the Salerno model \cite{Salerno} with competing on-site
and inter-site cubic terms \cite{Zaragoza}.

\begin{figure}[h]
\centering
\subfigure[$ $]{\includegraphics[scale=0.75]{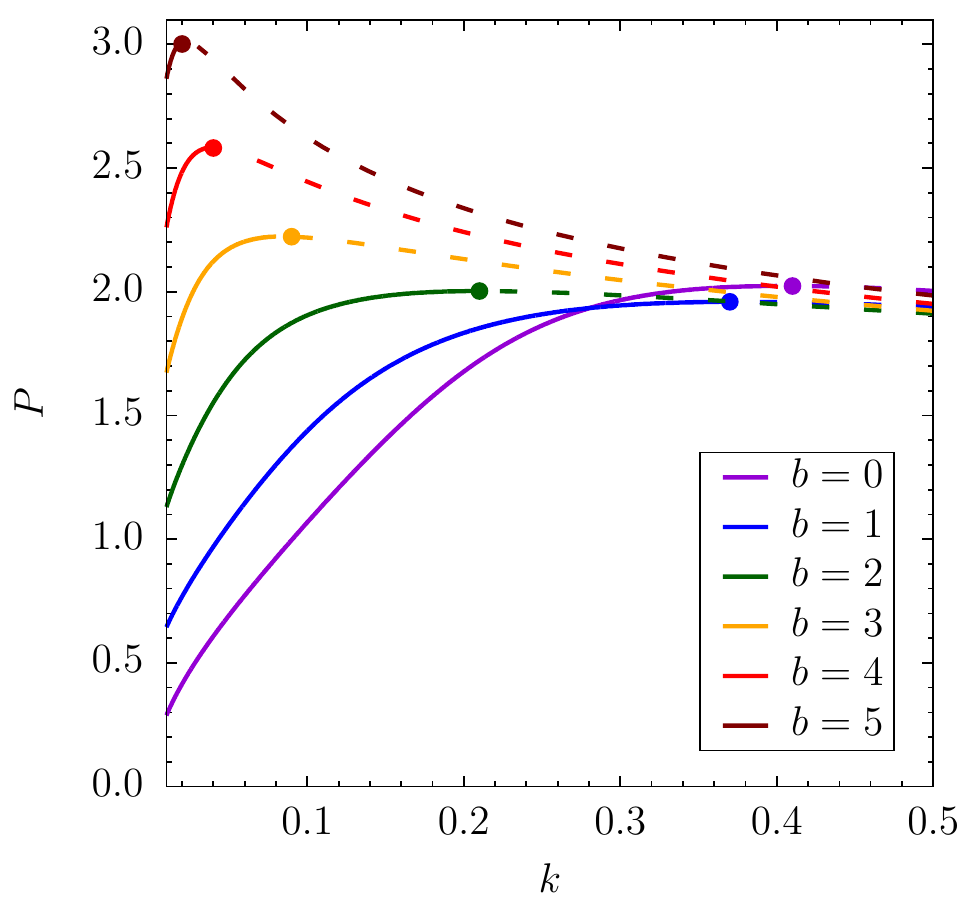}}
\subfigure[$
$]{\includegraphics[scale=0.75]{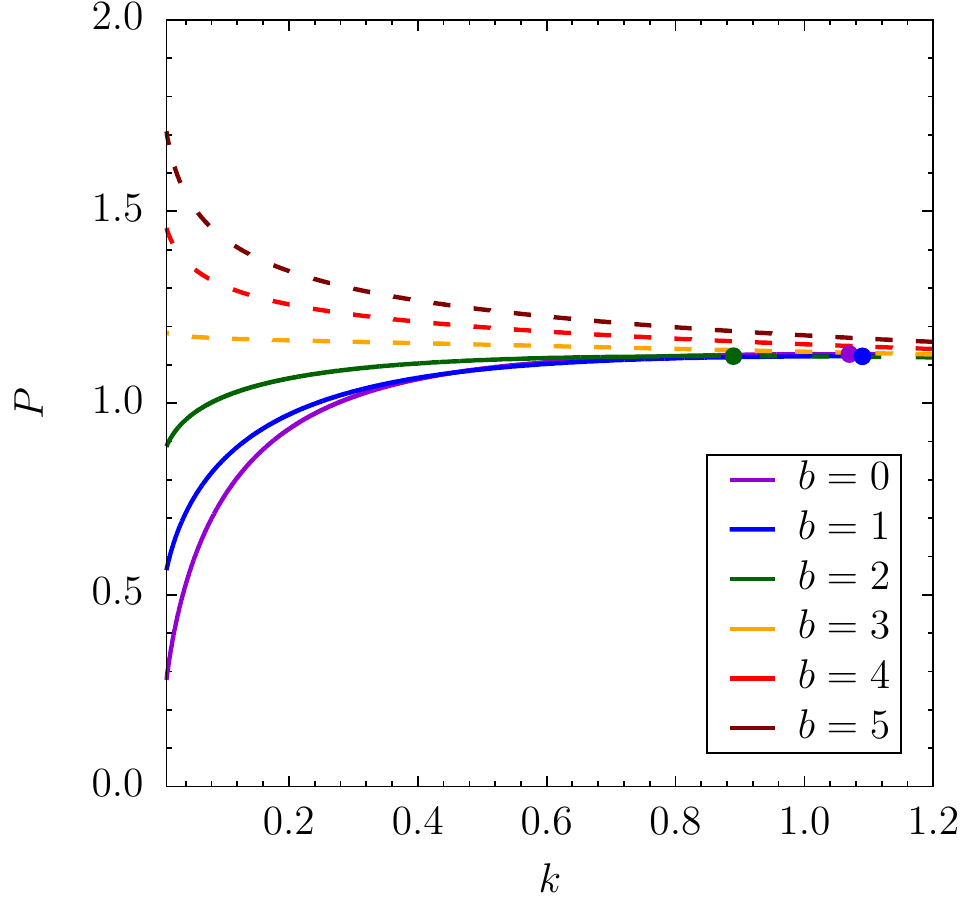}}
\caption{Propagation constant $k$, as a function of total power $P$, for
soliton families produced by the numerical solution of Eq. (\protect\ref%
{stationary}) with (a) $g_{5}=-1$ and (b) $g_{5}=+1$, which correspond,
respectively, to the self-defocusing and focusing quintic term, at different
values of the inverse-width parameter, $b$ in Eq. (\protect\ref{gaussian}) ($%
b=0$ corresponds to the spatially uniform nonlinearity). Solid and dashed
segments represent, severally, to stable and unstable solutions, according
to eigenvalues produced by the numerical solution of Eq. (\protect\ref%
{bdgsis3d}) (the VK criterion predicts exactly the same stability regions,
see the text). Bold dots designate the largest and smallest values of the
power, $P_{\max }$ and $P_{\min }$, see Eqs. (\protect\ref{<0}) and (\protect
\ref{>0}) below. In (a), all stable branches originate, at $k=0$, from
points $P=P_{\min }\left( g_{5}<0;b\right) $, and in (b) the unstable
branches (those corresponding to $b=3,4,5$) terminate at points with $k=0$
and $P=P_{\max }(g_{5}>0;b)$.}
\label{mu_n}
\end{figure}

Further, in Fig. \ref{mu_n} we present characteristics of numerically found
stationary soliton families, provided by the dependence of the propagation
constant, $k$, on the total power, $P$, as obtained from the numerical
solution of Eq. (\ref{stationary}) for $g_{5}=-1$ and $g_{5}=+1$, i.e.,
defocusing and focusing signs of the quintic term, with different values of
parameter $b$ which determines the inverse width of the
nonlinearity-localization region in Eq. (\ref{gaussian}). The stability of
the soliton families is also shown in Fig. \ref{mu_n}.

At $g_{5}<0$ (the defocusing sign of the quintic term) and each $b$, the
stable solitons exist in the interval of $P_{\min }\left( g_{5}<0;b\right)
<P<P_{\max }\left( g_{5}<0;b\right) $, with $P_{\min }\left(
g_{5}<0;b\right) $ and $P_{\max }\left( g_{5}<0;b\right) $ given (in an
approximate form) by Eq. (\ref{<0}) presented below. These smallest and
largest values of the total power are designated by bold dots in Fig. \ref%
{mu_n}.


In the limit of $k\rightarrow \infty $, the soliton solution, subject to
normalization condition (\ref{1}), becomes very narrow (hence it does not
depend on the value of $b$), with the shape dominated by the septimal term:%
\begin{equation}
\psi _{k\rightarrow \infty }(x)\approx \frac{k^{1/4}}{\left[ \cosh \left( 3%
\sqrt{2k}x\right) \right] ^{1/3}},  \label{lim}
\end{equation}
where the limit form of the $k(P)$ relation is
\begin{equation}
k\approx 15.2\cdot P^{-6}  \label{asympt-k}
\end{equation}
[numerical factors in Eqs. (\ref{lim}) (where the factor is very close to $1$%
) and (\ref{asympt-k}) are determined by the value of coefficient $%
\int_{-\infty }^{+\infty }\left( \mathrm{sech~}x\right) ^{2/3}dx\approx 4.2$%
]. The range of (very large) values of $k$ in which the asymptotic solution,
given by Eqs. (\ref{lim}) and (\ref{asympt-k}), is valid, is not shown in
Fig. \ref{mu_n} because the solutions are completely unstable in that case,
as Eq. (\ref{asympt-k}) does not meet the Vakhitov-Kolokolov (VK) criterion,
which is discussed in detail below (the instability is also corroborated by
numerical results).

It is worthy to note that the self-defocusing sign of the quintic term, $%
g_{5}<0$, provides for the expansion of the existence and stability limit to
much greater values of the total power, $P$. While this result is quite
natural in the case of the CQS\ nonlinearity, it is relevant to stress that
a formally similar combination of the focusing quintic and defocusing cubic
nonlinearities, in the absence of the septimal term, gives rise to
completely unstable solitons \cite{PhysicaD}. In the present case, the
stabilization is provided by the presence of the focusing term with the
lowest (cubic) nonlinearity, which is absent in the above-mentioned
cubic-quintic setting. As for the situation with the focusing quintic term, $%
g_{5}>0$, which is displayed in the Fig. \ref{mu_n}(b), it is relevant to
stress that each $k(P)$ curve features a stability segment.

Furthermore, it is noteworthy too that, taking larger $b$, i.e., a narrower
nonlinearity-localization region in Eq. (\ref{gaussian}), which is the most
essential feature of the present model, also helps to extend the stability
region of $P$. This effect is quite conspicuous, in the Fig. \ref{mu_n}(a),
for $g_{5}<0$, while for $g_{5}>0$ the stability interval expands, in the
Fig. \ref{mu_n}(b), up to some critical value of $b$, which is followed by
the transition to completely unstable soliton families. These observations
suggests to consider the limit form of the model with the modulation
represented by the delta-function, instead of the Gaussian in Eq. (\ref%
{gaussian}), which admits an exact analytical solution, as shown in the next
section.

\begin{figure}[h]
\centering
\includegraphics[scale=0.875]{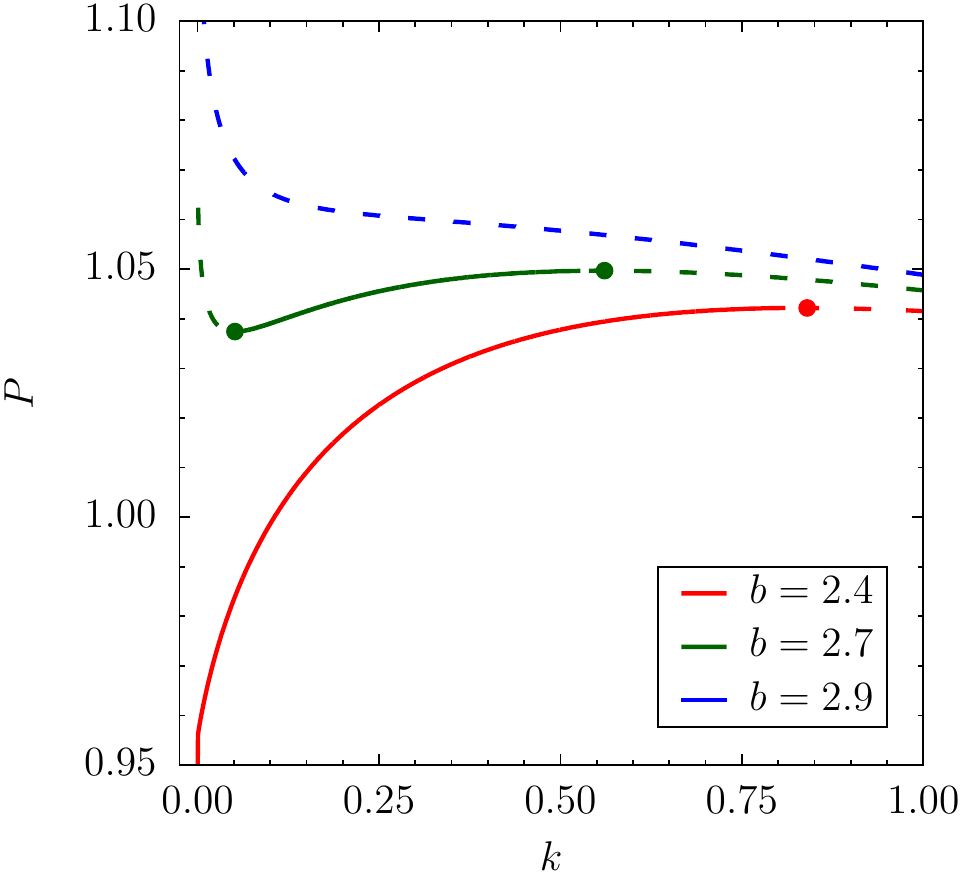}
\caption{The same as in Fig. \protect\ref{mu_n}, but for $g_{5}=1.5$ (the
self-focusing quintic term), in a narrow interval of values of $b$ around
the point of the transition from partly stable soliton families to
completely unstable ones.}
\label{mu_n_special}
\end{figure}

In the typical $k(P)$ plots displayed in Fig. \ref{mu_n}, each curve has a
single boundary between unstable and stable (if any) segments. In a more
special case, occurring at $g_{5}>0$ (the fully focusing nonlinearity) near
the transition between partly stable and fully unstable soliton families, $%
k(P)$ curves feature stable segments located between two unstable ones,
i.e., in a narrow interval of $P_{\min }<P<P_{\max }$, as can be seen in
Fig. \ref{mu_n_special}. In particular, the stability regions displayed in
the figure resemble results reported in Ref. \cite{gammal} for a liquid-gas
phase transition. \newline

\begin{figure}[h]
\centering
\includegraphics[scale=0.875]{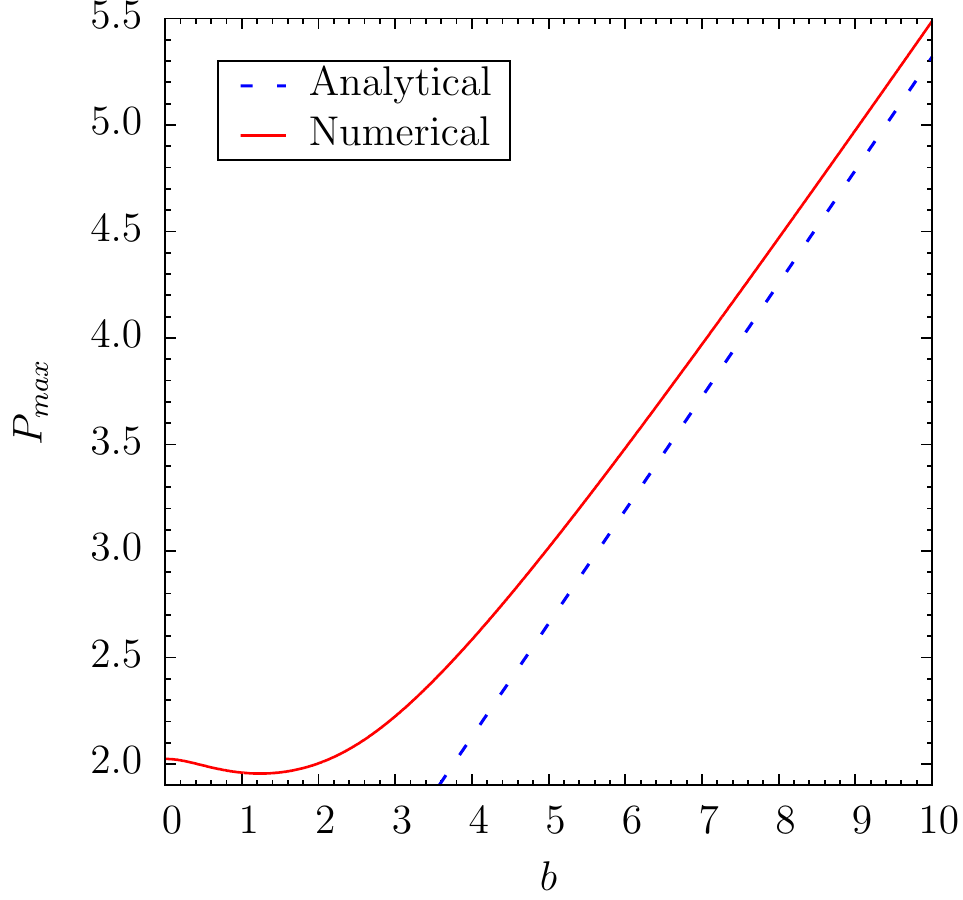}
\caption{The largest total power $P_{\max }$ for families of stable solitons
in the Fig. \protect\ref{mu_n}(a) as a function of $b$, for $g_{5}=-1$. The
red curve summarizes numerical results, while the blue one represents the
analytical result produced by the delta-functional limit, see Eq. (\protect
\ref{<0}).}
\label{asymptotic}
\end{figure}

The largest total power, $P_{\max }$, up to which stable solitons exist, is
an obviously important characteristic of the model with finite $b$ and $%
g_{5}<0$, as suggested by the Fig. \ref{mu_n}(a). This dependence is
displayed in Fig. \ref{asymptotic}, along with the analytical prediction
obtained in the delta-functional limit.

\begin{figure}[th]
\centering
\includegraphics[scale=0.85]{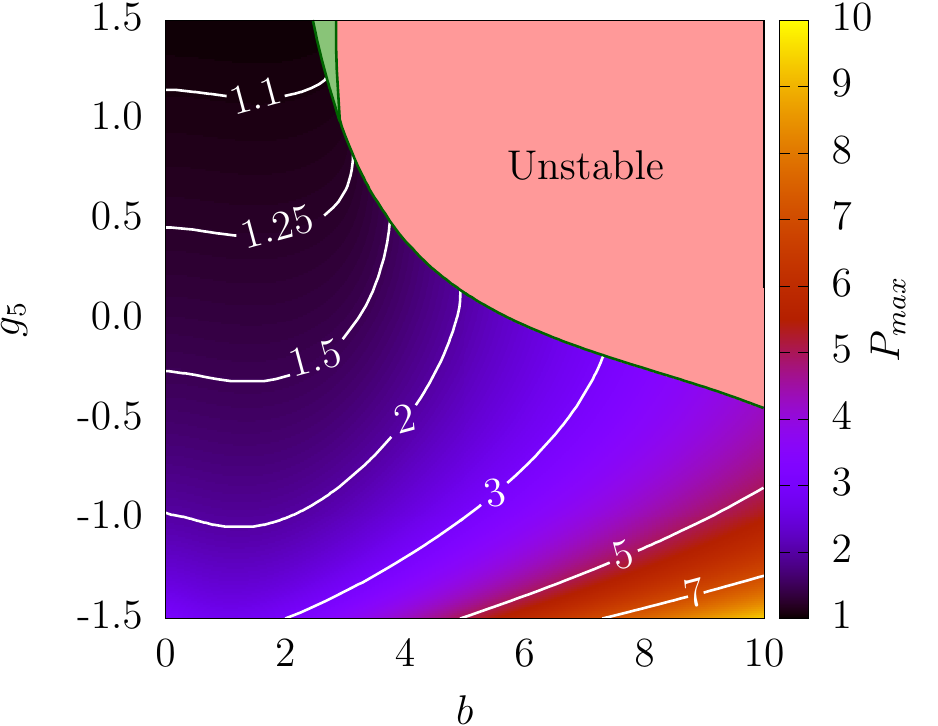}
\caption{The map of the largest total power of stable solitons, $P_{\max }$,
in the ($b$,$g_{5}$) parameter plane, as obtained from the solutions of Eqs.
(\protect\ref{time_dep})-(\protect\ref{=1}). $P_{\max }$ keeps a constant
value along each white curve. In the narrow green wedge, stable soliton
families feature essentially the same structure as demonstrated by the curve
for $b=2.7$ in Fig. \protect\ref{mu_n_special}.}
\label{map}
\end{figure}

Results of the systematic numerical analysis are summarized in Fig. \ref{map}%
, which displays $P_{\max }$ as a function of both control parameters, $%
g_{5} $ and $b$, in a broad area, $\left( 0\leq b\leq 10\right) \times
\left( -1.5\leq g_{5}\leq +1.5\right) $. This map shows, in the general
form, the trend which was stressed above in the connection to Figs.\ \ref%
{mu_n} and \ref{asymptotic}, \textit{viz}., that, at $g_{5}\leq 0$, the
increase of $b$, i.e., narrowing of the spatial localization of the
nonlinearity, leads to strong increase of the largest total power of the
stable solitons. This effect may find various applications, such as strong
confinement of powerful light beams in a narrow guiding channel by means of
the spatial modulation of the local nonlinearity, in the context of the
general topic of nonlinear light guiding \cite{Barcelona2}.

In addition to the accurate computation of the eigenvalues based on Eq. (\ref%
{bdgsis3d}), the stability of the soliton families can be determined by dint
of the VK criterion \cite{vakhitov,berge}, which gives the necessary
stability condition in the form of $dk/dN>0$ (in the general case, the VK
criterion is not sufficient for the stability, although in relatively simple
settings, such as the present ones, it may be sufficient). As is clearly
seen in Figs. \ref{mu_n} and \ref{mu_n_special}, in the present model the VK
criterion is, indeed, not only necessary but also sufficient for the
stability, as it precisely predicts the stability areas, identified by the
condition of $\mathrm{Im}\{\lambda \}=0$ [see Eq. (\ref{bdgsis3d})].

\begin{figure}[h!]
\centering
\subfigure{\includegraphics[scale=0.8]{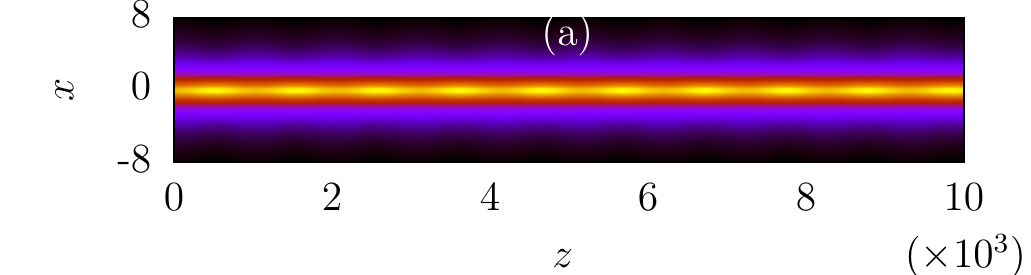}}\\ 
\subfigure{\includegraphics[scale=0.8]{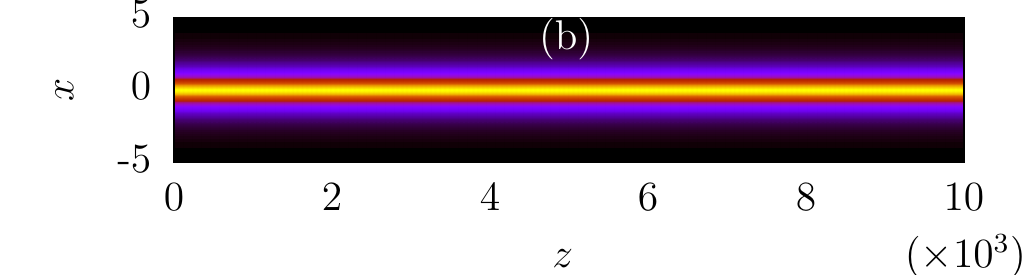}}\\ 
\subfigure{\includegraphics[scale=0.8]{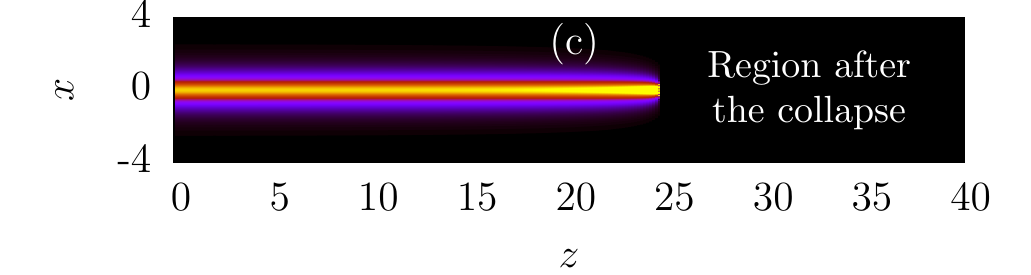}}
\caption{Top views of the simulated evolution of solutions for $g_{5}=1.5$
and $b=2.7$, with $k=0.04$, $P=1.038$ (a); $k=0.25$, $P=1.045$ (b); and $%
k=0.7$, $P=1.049$ (c).}
\label{real_time}
\end{figure}

\begin{figure}[h!]
\centering
\subfigure[]{\includegraphics[scale=0.9]{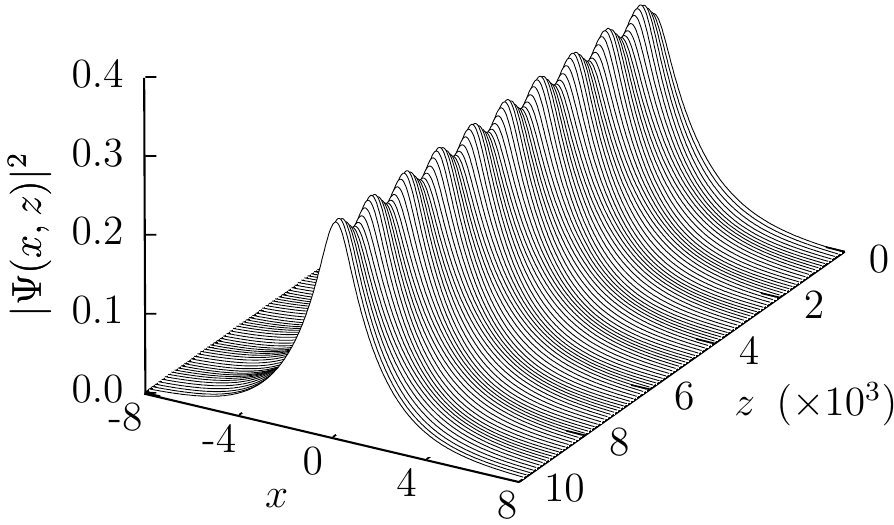}} \subfigure[]{%
\includegraphics[scale=0.9]{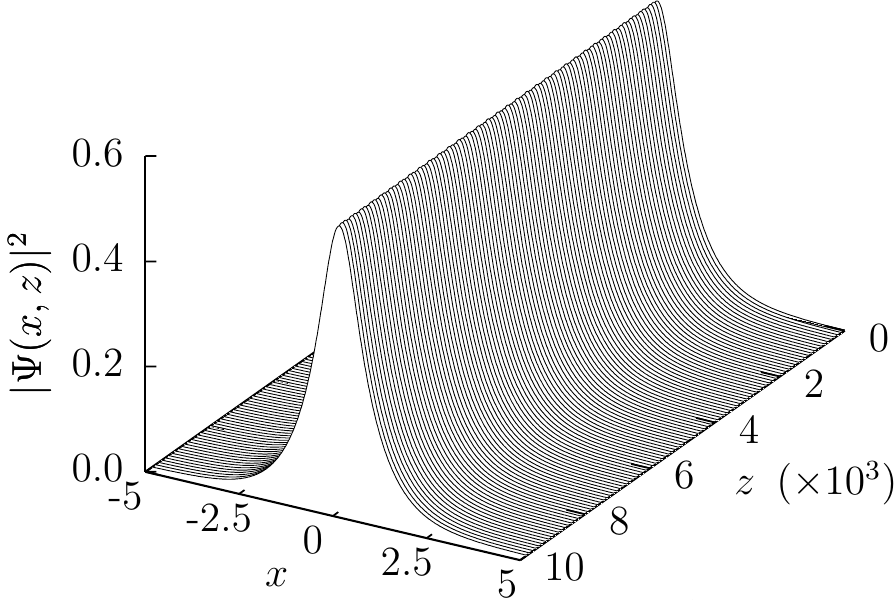}} \subfigure[]{%
\includegraphics[scale=1]{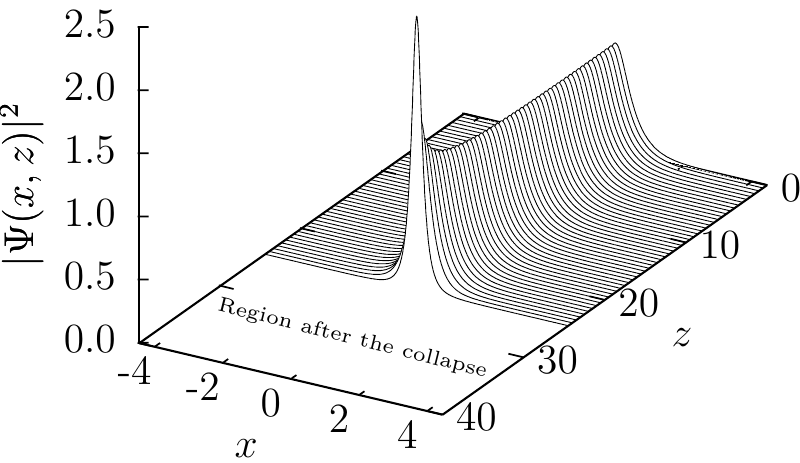}}
\caption{The evolution of an unstable soliton for $k=0.04$, $P=1.038$ (a),
stable soliton for $k=0.25$, $P=1.045$ (b), and unstable soliton for $k=0.7$%
, $P=1.049$ (c), which correspond, respectively, to panels (a), (b), and (c)
of Fig. \protect\ref{real_time}.}
\label{plot3d}
\end{figure}

The so predicted stability/instability of the solitons was tested by direct
simulations of their propagation, as shown in Figs. \ref{real_time} and \ref%
{plot3d}. In particular, at $g_{5}=1.5$, $b=2.7$, and $k=0.04$, $P=1.038$,
the soliton with broad tails [their size may be estimated as $\sim k^{-1/2}$%
, see Eq. (\ref{lim})], which, according to Fig. \ref{mu_n_special}, is
unstable, spontaneously transforms into a robust breather with a small
amplitude of internal vibrations, as shown in Figs. \ref{real_time}(a) and %
\ref{plot3d}(a). On the other hand, Figs. \ref{real_time}(c) and \ref{plot3d}%
(c) demonstrate that an unstable soliton with essentially shorter tails,
corresponding to $k=0.7$, $P=1.049$, suffers the collapse. Lastly, the
predicted stability of a typical soliton with $k=0.25$, $P=1.045$ is
confirmed by the simulations displayed in Figs. \ref{real_time}(b) and \ref%
{plot3d}(b).

\section{The analytical solution in the limit of the delta-functional
modulation profile}

As mentioned above, the model produces noteworthy results, such as the
possibility to support the transmission of tightly confined high-power
beams, at large values of $b$. In this case, the narrow Gaussian in Eq. (\ref%
{gaussian}) may be approximated by the delta-function:%
\begin{equation}
\exp \left( -b^{2}x^{2}/2\right) \approx \sqrt{2\pi }b^{-1}\delta (x).
\label{delta}
\end{equation}
Then, Eq. (\ref{stationary}) is replaced by

\begin{equation}
k\psi =\frac{1}{2}\frac{d^{2}\psi }{dx^{2}}+\frac{\sqrt{2\pi }}{b}\delta
(x)P\psi ^{3}\left( 1+g_{5}P\psi ^{2}+P^{2}\psi ^{4}\right) .
\label{stationary2}
\end{equation}
At $x\neq 0$, Eq. (\ref{stationary2}) amounts to a linear equation, $\psi
^{\prime \prime }=-2\mu \psi $, whose solution, satisfying normalization
condition (\ref{1}), is%
\begin{equation}
\psi (x)=\left( 2k\right) ^{1/4}\exp \left( -\sqrt{2k}|x|\right) .
\label{lambda}
\end{equation}
The integration of Eq. (\ref{stationary2}) in an infinitesimal vicinity of $%
x=0$ gives rise to the following boundary condition at $x=0$, assuming that $%
\psi (x)$ is an even function:%
\begin{equation}
\frac{1}{2} \frac{d\psi }{dx}{\LARGE |}_{x=0}=-\frac{\sqrt{2\pi }}{b}P\psi
^{3}\left( 1+g_{5}P\psi ^{2}+P^{2}\psi ^{4}\right) {\LARGE |}_{x=0}.
\label{x=0}
\end{equation}
The substitution of solution (\ref{lambda}) in Eq. (\ref{x=0}) yields a
quadratic equation for $\sqrt{k}$, which determines it as a function of $P$:%
\begin{equation}
2P^{3}k+g_{5}P^{2}\sqrt{2k}+\left( P-\frac{b}{\sqrt{2\pi }}\right) =0.
\label{quadr}
\end{equation}

As said above, an important characteristic of the soliton families is the
largest value $P_{\max }$ of the total power, up to which the solitons
exist. A straightforward analysis of Eq. (\ref{quadr}) yields, for $g_{5}>0$
(i.e., the self-focusing quintic term),%
\begin{equation}
P_{\max }\left( g_{5}>0;b\right) =b/\sqrt{2\pi }.  \label{>0}
\end{equation}
\ At $P<P_{\max }\left( g_{5}>0;b\right) $, there is a single soliton family
with $dk/dP<0$, which is definitely unstable, according to the VK criterion.
Note that $P_{\max }\left( g_{5}>0;b\right) $ given by Eq. (\ref{>0}) does
not depend on $g_{5}$ (as long as $g_{5}$ is positive). These analytical
findings, including the values of $P_{\max }\left( g_{5}>0;b\right) $ given
by Eq. (\ref{>0}), are in good agreement with the numerical counterparts
displayed in the Fig. \ref{mu_n}(b) for $b\geq 3$.

Further, the analysis of Eq. (\ref{quadr}) for $g_{5}<0$ demonstrates that
it generates two branches of the $k(P)$ dependence. The branch satisfying
the VK criterion, $dk/dP>0$, exists in a finite interval of values of the
power, under restriction $\left\vert g_{5}\right\vert <2$:%
\begin{eqnarray}
P_{\min }(g_{5} &<&0;b)=\frac{b}{\sqrt{2\pi }}<P<P_{\max }(g_{5}<0;b) \\
&=&\frac{b}{\sqrt{2\pi }}\left( 1-\frac{g_{5}^{2}}{4}\right) ^{-1}.
\label{<0}
\end{eqnarray}%
Equation (\ref{<0}) correctly explains an asymptotically linear dependence $%
P_{\max }(g_{5}=-1;b)$ at large $b$, which is clearly seen in Fig. \ref%
{asymptotic} . In particular, the slope of the asymptotic dependence is
exactly predicted by Eq. (\ref{<0}). Subsequent consideration of Eq. (\ref%
{quadr}) readily shows that $P_{\max }(g_{5}<0)=\infty $ at $g_{5}\leq -2$,
which actually implies that there remains a single $k(P)$ branch, with $%
dk/dP<0$, i.e., a completely unstable one.


\section{Conclusion}

In this work, we have addressed the question of how the recently
demonstrated stability area for solitons supported by the NLSE with the CQS
(cubic-quintic-septimal) nonlinearity is affected by the spatial confinement
of the nonlinearity, which may be realized in optical waveguides based on
colloids of nanoparticles. In particular, it was demonstrated that the
localization greatly increases the largest power of the stable guided beams,
which opens a path to transmit narrow high-power beams by means of the
nonlinear confinement. The stability of the spatial solitons in this model
is completely determined by the VK (Vakhitov-Kolokolov) criterion. The
results were obtained by means of the systematic numerical analysis, and
also in the analytical form, valid in the limit of the tight localization of
the nonlinearity.

It may be interesting to extend the analysis for a system of two parallel
nonlinearly confined waveguides, thus implementing a nonlinear coupler. A
challenging possibility is to explore a two-dimensional generalization of
the present model, that would imply a nonlinear core embedded in a linear
bulk medium (cf. a similar model with the self-focusing cubic-only
nonlinearity, considered in Ref. \cite{HS}).

\null
\null

\ack J.B.S. thanks Council of Scientific and Industrial Research (CSIR),
India, for the financial support. R.R. acknowledges DST (grant No.
SR/S2/HEP-26/2012), Council of Scientific and Industrial Research (CSIR),
India (grant 03(1323)/14/EMR-II dated 03.11.2014) and Department of Atomic
Energy - National Board of Higher Mathematics (DAE-NBHM), India (grant
2/48(21)/2014 /NBHM(R.P.)/R \& D II/15451 ) for financial support in the
form of Major Research Projects. H.F. and A.G. thank FAPESP and CNPq
(Brazil) for financial support. The work of B.A.M. is supported, in part, by
the joint program in physics between the National Science Foundation (U.S.)
and Binational Science Foundation (U.S.-Israel), through grant No. 2015616.

\end{document}